\documentclass{elsart}
\usepackage{natbib}
\usepackage{epsfig}
\begin{document}
\newcommand{\pst}{P_{\rm st}}
\newcommand{\pcr}{P_{\rm st}^{(C)}}
\newcommand{\gas}{\mu}
\newcommand{\gasat}{\gas_{\rm atomic}}
\newcommand{\gasact}{\gas_{\rm active}}
\newcommand{\gasinact}{\gas_{\rm inactive}}
\newcommand{\sfr}{{\rm SFR}}%  {\Sigma_{\rm SFR}}
\newcommand{\sig}{\Sigma_{\rm gas}}
\newcommand{\beq}{\begin{equation}}
\newcommand{\eeq}{\end{equation}}
\begin{frontmatter}
%\title{Star Formation Rates \\ and Propagating Stochastic
% Star Formation} \\ in Spiral Galaxies}
\title{Comment Regarding  the \\ Functional Form of the Schmidt Law}
\author{Peter Todd Williams}

\begin{abstract}
Star formation rates on the galactic scale are described phenomenologically by two distinct
relationships, as emphasized
recently by Elmegreen (2002). The first of these is the Schmidt law, which is a power-law
relation between the star formation rate $SFR$ and the column density $\Sigma$.
The other relationship is that there is a cutoff in the gas density below which
star formation shuts off.

The purpose of this paper is to argue that 1) these two relationships can be accommodated by a single functional form of the
Schmidt law, and 2) this functional form is motivated by the hypothesis that star formation is a critical
phenomenon, and that as a corollary, 3) the existence of a sharp cutoff may thus be an emergent property of galaxies,
as was argued by Seiden (1983), as opposed
to the classical view that this cutoff is due to an instability criterion.
\end{abstract}
\end{frontmatter}

%%%%%%%%%%%%%%%%%%%%%%%%%%%%%%%%%%%%%%%%%%%%%%%%%%%%%%%%%%%%%%%%%%%%%%%%%%%%%%%%
\section{Introduction}
%In this paper, we discuss some theoretical connections between star-formation rates as expressed by
%the Schmidt law with a lower cutoff, and simple stochastic self-propagating star-formation (SSPSF) models.

The Schmidt law 
is a power-law relationship between the star formation rate $\sfr$ and the gas column density $\Sigma$, introduced
by Schmidt (1959), namely
\beq
\sfr \propto \Sigma^n
\label{eq:schmidt}
\eeq
for some exponent $n$. Schmidt estimated $n$ to be close to 2; modern measurements of the Schmidt index tend
to give lower values for $n$, such as 1.4 (Kennicutt, 1998; Heyer et al., 2004),
although the index depends upon
what material is included (neutral gas, ionized gas, stars) in $\Sigma$. Note that Wong \& Blitz (2002) indicate that the most
reliable tracer is the surface density of molecular hydrogen; even so, there is still significant variance 
in inferred values of $n$, in part due to observational and modeling uncertainties, but presumably in part due as well
to real variation in Nature.

The other relationship is that there is a cutoff in the gas density below which star formation shuts off. Again, \cite{elm2002}
emphasizes that this cutoff is either interpreted phenomenologically as a fixed gas density, or a density determined by a stability criterion
(e.g.~the Toomre~Q criterion). The history of this cutoff is explored nicely in \cite{huelba1998}.

The basic Schmidt Law of eq.~({\ref{eq:schmidt}}) has many variants, which we will not describe here. 
The consistent notion is that, averaged over some suitably
large lengthscale, the star formation rate in a galaxy increases with the gas density, and that it does so as a power law. 
%Let us point out that a power
%law is simply a straight line living in log-log coordinates, so that to the
%extent that astronomers deal extensively with logarithms, a power law is a natural functional form to fit things to. 
What is surprising is how well this
 prescription seems to work, over a wide range of densities, from spiral galaxies up through starburst galaxies. Given the simplicity of eq.({\ref{eq:schmidt}}),
it is perhaps not too surprising that a wide variety of physical mechanisms have been suggested as the root of this behavior. Furthermore, various modifications
to this functional form have been suggested. The original prescription, as well as subsequent modifications, all share the property that the 
power law is centered on the origin. In other words, the two quantities of concern are connected by a relationship of the form
\beq
(x-x_0) = A (y-y_0)^\beta
\eeq
with the particular choice $x_0 = 0$ and $y_0 = 0$.

The problem with this is that it is at odds with the second observational fact described above, namely the existence of a cutoff. The result is
that, as seen in observations, star formation falls off of the power-law curve as one approaches the cutoff from above. This drop-off of the power-law
is also seen in simulations that show a cutoff (Kravtsov 2003).

The existence of a sharp cutoff in star formation below a critical surface density in spiral galaxies is shown in the data of Kennicutt (1989) and
of Martin \& Kennicutt (2001), among others  (although note that Pohlen et al.~(2002) argue against a sharp cutoff).
Dynamical instability (Toomre, 1964; Quirk, 1972; Wang \& Silk, 1994; Pandey \& van~de~Bruck, 1999) of some sort may explain this cutoff, as argued by Kennicutt (1989) and others.
On the other hand, the results of Wong \& Blitz (2002) and Boissier et al.~(2004) do not offer as strong support for the stability criterion interpretation
of the cutoff as might be hoped. 
In addition, \cite{huelba1998}
point out that it is not clear that the instabilities used to justify theoretically the existence of a cutoff critical density $\Sigma_0$ should 
actually apply directly to the star-formation process. Furthermore, recent star formation may actually have taken place in spirals beyond the nominal
cutoff radius, so that the observed cutoff may be to a degree an issue of observational sensitivity (Ferguson et al. 1998).

It is interesting to note that star formation behaving as a power-law with lower cutoff is also seen in certain extremely simple models of galactic-scale
star formation, namely self-propagating star formation (SPSF) and stochastic self-propagating star formation (SSPSF). Such models will also show a small
but significant degree of star formation below the cutoff, depending on the parameters of the simulation. On the other hand, it is not at
all clear that star-formation on galactic scales proceeds by self-propagation. For example, one current view is that supernovae (SN) are effective pumps
of supersonic turbulence in potential star-forming clouds, and that such turbulence provides the dominant means of support against Jeans collapse and
star formation (Mac~Low \& Klessen, 2004), implying that star formation in one locale is hindered rather than helped by recent star formation in nearby locales.

Here we do not advocate SPSF or SSPSF one way or another. However, we do take some lessons from the perspectives these models provide, such as the point of view that
 the star formation cutoff may an emergent property of the system, and might not be reflective of a classical dynamical instability. Instead, as
has been argued previously (see below), star formation and cutoff may be similar to critical phenomena more familiar to condensed-matter physicists.
In such critical phenomena, the cutoff is incorporated into the power law, as shown below.

\subsection{SPSF and SSPSF models}
As motivation to the points of view alluded to above and discussed further below, we now outline the basic considerations of
stochastic self-propagating star formation models. It should well be borne in mind, however, that the phenomenon of a critical point
in global star formation rates does not necessarily imply that star formation is indeed  a self-propagating phenomenon; rather, this
specific instance of critical behavior in global star formation models is simply offered as motivation for our modified functional form.

The self-propagating star-formation model (SPSF) was originally proposed by \cite{MuAr:1976} as a simple model that offered an  explanation of spiral
patterns in disk galaxies as the result of two processes, namely the presence of large-scale shear in disk galaxies, and the
hypothesis that star formation in one location induces star formation in neighboring regions. \cite{GeSe:1978} introduced the stochastic
self-propagating star-formation model (SSPSF model), which  modified the model of Mueller \& Arnett so that
 the inducement of star formation is probabilistic in nature. 
The SPSF/SSPSF  interpretation of spiral arms may be particularly apropos in the case of
so-called ``flocculent'' galaxies, which appear rather spatially disordered, as opposed to ``grand-design'' spirals in which the spiral arms are globally
coherent and appear to be dynamical in nature.
Both deterministic and stochastic SPSF models have been explored
more or less continuously in the literature since the publication of these two papers. \cite{ScSe:1986} provide an early review of
the basics of SPSF models.

One of the aspects of the SSPSF model that was noted early on by \cite{SSG:1979} is that the equilibrium star formation rate as a function
of the control parameter $\pst$ (which parametrizes the probability that star formation is induced in a given cell given that there
is star formation in a neighboring cell) exhibits a phase transition:
%  \footnote{The process has been referred to as percolation. Strictly speaking, 
%  percolation (e.g.,~bond percolation) is a time-independent process, so it
%  is not altogether clear to us whether a formal analogy holds. We therefore restrict ourselves to describing star formation in SSPSF models as a {\em critical} phenomenon.}:
The onset of star formation in SSPSF models as a function of $\pst$ is a critical phenomenon,
in which the star formation rate $\sfr$ is an order parameter (one of several) and the probability $\pst$ is the control parameter. Below some
critical probability $\pcr$ the star formation rate $\sfr$ is very low, becoming essentially zero far below the critical point. As one nears the critical point $\pcr$ from below,
the system undergoes larger and larger fluctuations; these fluctuations may also be seen above the critical point in the case that
the computational domain is very small. For example, the star formation may exhibit large bursts between protracted latency periods
with no star formation, as discussed in the theory of dwarf galaxies discussed in \cite{GSS:1980} and subsequent papers and by \cite{hir:2000}. 
These bursts are initiated by random star-formation seeding events that are controlled
by a second parameter, the spontaneous seeding rate, that is of little importance above the critical point so long as it is sufficiently small.

Far above the critical point, star formation is more or less constant in time.
Approaching the critical point from above, star formation as a function of time becomes more and more erratic, and in the absence of
seeding events it shuts off entirely below the critical point. Most importantly, the average (as a function of time and space) star formation
rate $< \sfr > $, as noted by \cite{SSG:1979}, behaves as
\beq
\sfr = \cases{A (\pst-\pcr)^\beta
 &if $\pst\ge
\pcr$;\cr 0 &otherwise,\cr}
\label{eq:critical}
\eeq
which is characteristic behavior for a critical phenomenon (see fig.~1a). Other properties of the system (such as the mean size of star-formation clusters, the variance
in the star formation rate, etc)
also depend upon $\pst$ with the same functional form, albeit with different critical indices. 

These relationships hold best near the critical point, but computational restrictions limit one
from reproducing this behavior arbitrarily close to that point:
Any real computation is necessarily limited to a finite number of cells, and this affects the dynamics of the system
very close to the critical point, because correlation lengths diverge there. Furthermore, the small but nonzero
rate of spontaneous star-formation that is an ingredient of most SSPSF models blunts the cusp in the phase transition at the
critical point, just as, for example, the random-field Ising model departs from canonical critical behavior close to the critical point
in that system. Also, behavior of the system departs from relationship (\ref{eq:critical}) far above the critical point; eventually
the system saturates and $\sfr$ becomes less dependent upon $\pst$. However, let us keep in mind that, as noted by \cite{SSG:1979} and \cite{ScSe:1986},
the  spiral patterns that were the original motivation for the SSPSF model are best produced with $\pst$ just above
the critical value. 

As a demonstration of the functional form (\ref{eq:critical}), in figure~(A.1) % figure~(\ref{fig:runfit}) 
we show $SFR$ as a function of $\pst$ in a very simple SSPSF code
we have written. This code implements SSPSF in a fixed 2-D (400 x 400) Cartesian lattice with periodic boundary conditions. Simulations are
followed for 10000 time steps, and the initial 1000 time steps are discarded from the analysis to remove transient behavior.
Cells are in one of four states:
ready for star formation, forming stars, supernova, and dormant. On each timestep, for a cell that is ``ready,'' the number of nearby
(above, below, left and right) supernova are added up, and multiplied by the parameter $\pst$. If a random number drawn from the uniform
distribution on [0,1] is less than this number, then star formation is initiated on the next step. The cell then goes ``supernova'' on
the subsequent step, then it is ``dormant,'' and finally it returns to being ``ready.''
There is random spontaneous seeding of star formation inserted at a very low rate (roughly one per 20 time steps for the entire mesh in this case).
No attempt is made to account for the gas consumed by star formation.
This is an extremely simplified version of SSPSF, used simply to illustrate the form in eq.~(\ref{eq:critical}) for this system, as shown also in
\cite{SSG:1979}.
Here we find the exponent $\beta$ to be $0.52$, close to the mean-field value of $1/2$, when we fit over the range $0.0 \le P \le 0.50$. At higher
values of $P$ the star formation rate begins to saturate.

Just as this cutoff in SSPSF models is called a phase transition, so too is the cutoff in observations of star formation referred to as a phase transition.

\subsection{Hypothesis}

Our hypothesis, based on the above notions, then, is that star formation should be written as a power-law function not of the
gas density, but of the difference between the gas density and the critical (or threshold) gas density, in analogy to the behavior
of critical phenomena:
\beq
\sfr \propto \cases{ (\Sigma-\Sigma_0)^n
                          & if $\Sigma \ge \Sigma_0$;\cr 0 &otherwise,\cr}
\label{eq:schmidt2}
\eeq
Here, the density $\Sigma$ plays the role of the control parameter. 
%Although it seems reasonable to assume that $\pst$ in SSPSF models is
%related to the areal density $\Sigma$, the relationship is not clear. Offhand, therefore, 
%it is not clear whether and how the exponent $n$ should be related to
%the exponent $\beta$. 
We do not address here the extensive literature on the various universality classes in the
theory of critical phenomena, and we make no attempt to connect the exponent $\beta$ from SSPSF models to the
Schmidt index $n$. Rather, the point is merely that, within the theory of critical phenomena, power laws are always expressed about the critical
point, and so insofar as star formation may be thought of as a type of phase transition, it seems natural to incorporate the
cutoff into the Schmidt law in the form of eq.~(\ref{eq:schmidt2}).

%With this caveat in mind, it may be interesting to note the following coincidence: Let us suppose that the probability that star formation
%in one location induces star formation in neighboring regions (in the SSPSF model) is proportional to the density $\Sigma$ in the ``real world.''
%The star formation rate we show in (fig. XXX) is the time-averaged fraction of cells undergoing star formation in our simulation, which we have
%denoted ``$\sigma$''. To interpret this as a rate $R$ of star formation in units of mass per area per unit time, we must multiply it by the density
%(and some efficiency factor). We then obtain $R \propto (\Sigma - \Sigma_0)^{1 + \beta}$. If we take the value of the critical exponent $\beta$
%from our fit above, we obtain a Schmidt exponent of $1.40$, which is basically the current accepted value. 

Note that both eq.~(\ref{eq:schmidt}) with cutoff and eq.~(\ref{eq:schmidt2}) have the same number of free parameters, namely three:
An amplitude (the proportionality constant, not written), a power-law index, and a cutoff.

\section{Data Analysis}
To test the relative merit of  eq.~(\ref{eq:schmidt}) with cutoff and eq.~(\ref{eq:schmidt2}), we fit two sets of data to these two functions.
The $\chi^2$ for these fits is found by assuming that the uncertainties in the logarithm of the physical quantity on the abscissa and the
ordinate are equal and constant for all data points. This greatly simplifies the fitting process described below.
Rather than tabulate the $\chi^2$ for the fits, we give  the
ratio of the $\chi^2$ for the best fits for the two functional forms,
since this ratio does not depend upon the actual value for the
uncertainties of the measurements. 

To fit the ordinary Schmidt law, we use established methods for fitting to a straight line with uncertainty in both abscissa and ordinate,
with the following modifications: A line is drawn perpendicular (in log-log space) to the fitted line at the cutoff point. Points to the left of this line contribute
to the $\chi^2$ of the fit by an amount proportional to the distance of these points from the cutoff point. We then minimize $\chi^2$ using
a downhill simplex method (\cite{NUMREC}).

To fit the modified Schmidt law, we estimate a relative $\chi^2$ by finding the closest straight-line distance on the
log-log plot between each point and the fitted curve. This requires a function minimization for each data point.
Again, we minimize $\chi^2$ using a downhill simplex method.

The first set is the data compiled by Hunter et al.~(1998), which includes ${\rm H}\alpha$ surface brightness as a proxy for star formation rate and total (atomic and molecular) 
gas density inferred by linear proportionality
to measured ${\rm HI}$ for a range of
irregular galaxies.
%This data covers a wide range of gas densities, and so to help discriminate between our two functional forms, we restrict our
%attention here to star formation in the outer half-radius of the star formation region in these galaxies. 
We throw out two galaxies
from this data set. One, DDO~105, does not extend in radius out past the star formation cutoff. The other, DDO~155, has few data points in
the star-forming region.

These are dwarf irregular galaxies, and star formation in such galaxies differs from star formation in spirals in many
respects. Indeed, dwarf galaxies do not show a strong correlation between ${\rm H}\alpha$ and ${\rm H\,I}$ (Brosch et al. 1998). Following
Wong \& Blitz (2002), it may be that much tighter correlations would be apparent here if data were available for molecular hydrogen (${\rm H_2}$)
instead of atomic hydrogen.

The data as plotted here consists of ${\rm H}\alpha$ surface brightness in ${\rm erg}/{\rm sec}/{\rm pc^2}$
as a measure of star formation rate versus gas column density quantified in $M_\odot / {\rm pc}^2$.
Fitting parameters $a, n,$ and $c$ correspond to the formulas
\beq
\log_{10}(\Sigma({\rm H}\alpha)) = \cases{a  + n\,\log_{10}(\Sigma_{\rm gas})
 &if $\Sigma_{\rm gas}\ge \Sigma_{\rm gas}^{(0)}$;\cr -\infty &otherwise,\cr}
\label{eq:fit1}
\eeq
where $c = \log_{10}(\Sigma_{\rm gas})$, in the case of a fit to a standard-form Schmidt law, and
\beq
\log_{10}(\Sigma({\rm H}\alpha)) = \cases{a  + n\,\log_{10}(\Sigma_{\rm gas} - \Sigma_{\rm gas}^{(0)})
 &if $\Sigma_{\rm gas}\ge \Sigma_{\rm gas}^{(0)}$;\cr -\infty &otherwise,\cr}
\label{eq:fit2}
\eeq
in the case of a fit to the modified Schmidt law of the form of eq.~(\ref{eq:schmidt2}). In particular, the quantity
$n$ here is thus the usual Schmidt exponent.

Table~1 shows that the results are equivocal here. For some galaxies, a simple power-law fit with cutoff is a better fit, whereas for others
our modified power-law form appears to be a better fit. Clearly, for some galaxies, both equations offer quite poor fits.

\begin{table}
\caption{Relative Merit of Fits for Data of Hunter et al.~(1998); see text for definitions of $a$, $n$, and $c$.}
\begin{tabular}{l c c c c c c c}
       &        &  standard    &             &                 &  modified      &          &           \\
\cline{2-4}  \cline{5-7} 
object & a &  n   &  c    &a  &  n   &  c  &   $\chi^2_{\rm mod} / \chi^2_{\rm std}$ \\
%\cline{1-2}
%\hline Hunter & Elmegreen: & & & & & & \\
%  ddo105        -            -         -           -           -            -           -           -             -              
\hline
 DDO~154    &  26.5926 & 5.4241 & 0.7776 & 30.9764 & 1.5926 & 0.7546 & 1.6148 \\
%  ddo155        -            -         -           -           -            -           -           -             -              
 DDO~168    &  29.8522 &  1.6172 & 0.5434  &  31.201   &  0.6682  &  0.6965  &     0.184 \\ 
 DDO~50    &  26.1368 & 5.5392 & 0.8159 & 30.2791 & 2.4847 & 0.7864 & 1.0009 \\
 IC~1613  &  31.009 & 7.8061 & 0.4186 & 17.567 & 24.0863 & 0.4779   &   0.927 \\
 Sex~A    &   22.1768 & 10.8149 & 0.5042 & 26.4300 & 7.5003 & 0.4004 & 1.0246 \\
\end{tabular}
\end{table}

For our second analysis, we choose to reconsider the data of \cite{Ken:1989}. In particular, we
analyze the data taken from his figure~8 on page~694 of that paper. These data are measured ${\rm H}\alpha$ surface brightness versus published
hydrogen (${\rm HI} + {\rm H}_2$) gas densities, and are not available
in \cite{Ken:1989} in tabular form. Therefore, we simply recreate the original data by measuring the positions of each of the data
points on the figure. A similar plot is also found in \cite{Ken:1998}, however this plot does not clearly have individual data points marked,
and we were unsuccessful at attempting to recreate the data plotted there.

\begin{table}
\caption{Relative Merit of Fits for Data of Kennicutt (1989)}
\begin{tabular}{l c c c c c c c}
       &        &  standard    &             &                 &  modified      &          &           \\
\cline{2-4}  \cline{5-7} 
object & a &  n   &  c    &a  &  n   &  c  &   $\chi^2_{\rm mod} / \chi^2_{\rm std}$ \\
%\cline{1-2}
%  Kennicutt: 
\hline
 all    &      29.59   &    2.352  &    0.3255    &        31.36    &   1.085  &    0.5301   &          0.606 \\
  M101   &      29.28  &     2.944 &     0.2533   &        30.45   &    1.952  &    0.2422   &         0.962  \\
  M51    &      29.20  &     2.297 &     0.3150    &       30.25   &    1.639  &    0.3797   &          0.771 \\
  NGC~4254  &     30.84  &     1.372  &    0.6207   &       31.10   &    1.224  &    0.2039    &         1.061 \\
  NGC~4303  &    31.14   &    1.277  &    0.4325   &       32.11   &    0.670  &    0.6762    &         0.647 \\
  NGC~4321  &    30.92   &    1.336  &   -0.3723   &        31.68   &    0.791  &    0.2028   &         0.178 \\
  NGC~4535  &    30.03   &    2.124  &    0.2068  &        31.38   &    1.072  &    0.4679   &          0.476 \\
  NGC~6946  &    28.67   &    3.027  &    0.6132   &       31.97   &    1.435  &    0.6533  &           0.287 \\%
\end{tabular}
\end{table}

Plots of these fits are included in the appendix.

It can be seen from these fits that the Schmidt exponent is unusually high in several cases. This can be attributed in part to fitting unabashedly
the entire curve, including
downturn that is the subject of this paper. Note that the exponents for the modified Schmidt law are systematically lower, as would be expected.

Note as well that the results here are not nearly so equivocal as before; in particular, the relative merit of fit to the modified SFR law is
markedly better here than for the standard form, as demonstrated by the $\chi^2$ ratios for the two fits.

\section{Conclusions}
We have offered here an interpretation of the star formation cutoff as an emergent property of galactic systems, in the vein of SSPSF models,
following on the work of \cite{sei1983}. This motivates us to consider a modification to the Schmidt law, such that star formation is written
as a power law around the critical density, in analogy to the functional form of quantities near a critical point. This new functional form
captures the downturn in star formation rates at gas densities near the cutoff point. We find that this functional
form is a better fit to the data for spiral galaxies, but for irregular galaxies the results are ambiguous.

While this arguably provides some evidence in support of the hypothesis of \cite{sei1983}, we note that there are several other interpretations
of the downturn of star formation rate near the cutoff.

In particular, some other interpretations include:
1) Star formation occurs through a global instability with a critical density. Given a quadratic form for the dispersion relation,
the growth rate of the instability, as a function of density, is proportional to $\sqrt{\sig-\Sigma_0}$. If star formation is
proportional to the density times the growth rate of the instability, we obtain $\sfr \propto \sig \sqrt{\sig - \Sigma_0}$, which
asymptotes to an exponent of $1.5$ and, moreover, approaches zero nicely from above.
(Note that this is one of a class of relations of the form $\sfr \propto \sig^\beta (\sig - \Sigma_0)^{n-\beta}$.)
2) The star formation rate is intimately connected with the high-density tail of the density PDF of supersonic turbulence in the ISM,
   as suggested by Kravtsov (2003).
   The functional dependence of star formation on density is reflective of this tail, which has a power-law form with an upper cutoff,
   resulting in a power-law dependence of star formation on gas density with a lower cutoff. In fact, we show in the appendix that
   our modified functional form appears (by eye) to be a much better fit to the simulations of Kravtsov (2003) than the standard
Schmidt law. This cannot be due entirely to feedback processes (as in SSPSF models) because Kravtsov (2003) finds a downturn and cutoff even
 when feedback is turned off.
3) Star formation really does have a power-law dependence with instantaneous cutoff, and the appearance of a steepening in the Schmidt index near the cutoff is simply
   due to observational mixing of sub-critical and super-critical regions. 

Thus, our functional form does not discriminate between the cutoff as an emergent property of the system, and other hypothesis regarding the
origin of the cutoff. However, we suggest that the critical phenomenon interpretation predicts that a variety of different quantities depend upon the parameter $\pst$ in
the form of eq.~(\ref{eq:critical}), and comparison of the behavior of these quantities with respect to parameters such as the gas density $\Sigma$ may
offer further evidence in support of the critical-point hypothesis for the origin of a cutoff. The deviations from a sharp cutoff in star formation
that have been observed in extreme outer regions of disk galaxies as noted above (Ferguson et al. 1998) 
actually may be evidence for, rather than against, a critical phenomenon
interpretation, as sytems exhibiting critical phenomena deviate from the form eq.~(\ref{eq:critical}) both if the system is subject to noise and if the
system is small enough not to approach the thermodynamic limit. Indeed, as the number of star formation sites in galaxies is nowhere near large enough to approach the nominal
thermodynamic limit, unlike laboratory systems ($\sim 10^{23}$ particles), it should not be surprising that the cutoff is not so sharp. However, accomodation of this
fact clearly demands modification of eq.~(\ref{eq:critical}) through the introduction of another parameter to account for such finite-size effects, and is 
beyond the scope of the work presented here.

%%%%%%%%%%%%%%%%%%%%%%%%%%%%%%%%%%%%%%%%%%%%%%%%%%%%%%%%%%%%%%%%%%%%%%%
%\begin{deluxetable}
%\tablewidth{Opt}
%\tablecolumns{}
%\end{deluxetable}
%%%%%%%%%%%%%%%%%%%%%%%%%%%%%%%%%%%%%%%%%%%%%%%%%%%%%%%%%%%%%%%%%%%%%%%
%\acknowledgments
%%%%%%%%%%%%%%%%%%%%%%%%%%%%%%%%%%%%%%%%%%%%%%%%%%%%%%%%%%%%%%%%%%%%%%%
{\it
Much of this work was performed while the author was a graduate student at the
University of Texas at Austin.
The author thus wishes to thank the theory group in Austin for their support. He also wishes to thank
D.~Hunter and A.~Kravtsov for kindly providing results of their observations and simulations, respectively.
%This work was supported in part by the E.D.D. of the State of California.
}

%\bibliographystyle{plain}
%\bibliography{schmidt}

\begin{appendix}
\section{Fits to Data}
Here we show fits to data and simulations. First, we show results from our simple
SSPSF implementation (fig. 1a).
Next, we show fits to the data compiled by Hunter et al (1998) for irregular galaxies (fig. 2, c--g) We show fits to ${\rm H}\alpha$ surface
brightness for concentric annuli, used as a proxy for star formation rate.
The abscissa is total hydrogen  (${\rm HI} + {\rm H}_2$) gas density, inferred from measured ${\rm HI}$ emission.
We then show fits to the data collected in Kennicutt (1989) for Sc spiral galaxies (fig. 2 and 3, h--n). 
Here the abscissa is the total hydrogen (${\rm HI} + {\rm H}_2$) gas density, where atomic and molecular
densities were taken from previous studies in the literature.
Finally, we show fits to the simulated data of Kravtsov (2003) (fig. 1b). Note that in this final
case, the relative $\chi^2$ for the goodness of fit, i.e.~$\chi^2_{\rm mod} / \chi^2_{\rm std}$,
is just barely less than one, namely it is $0.956$. However, this measure fails to capture the
better ``Chi-by-eye'' feature of the points that they are {\em curved}, and that
this curve is captured by the modified functional form.

\begin{figure}[htp]
\centering
\epsfxsize=6in
%\epsfbox{/Users/petwil/projects/schmidt/data/figure_1.eps}
\epsfbox{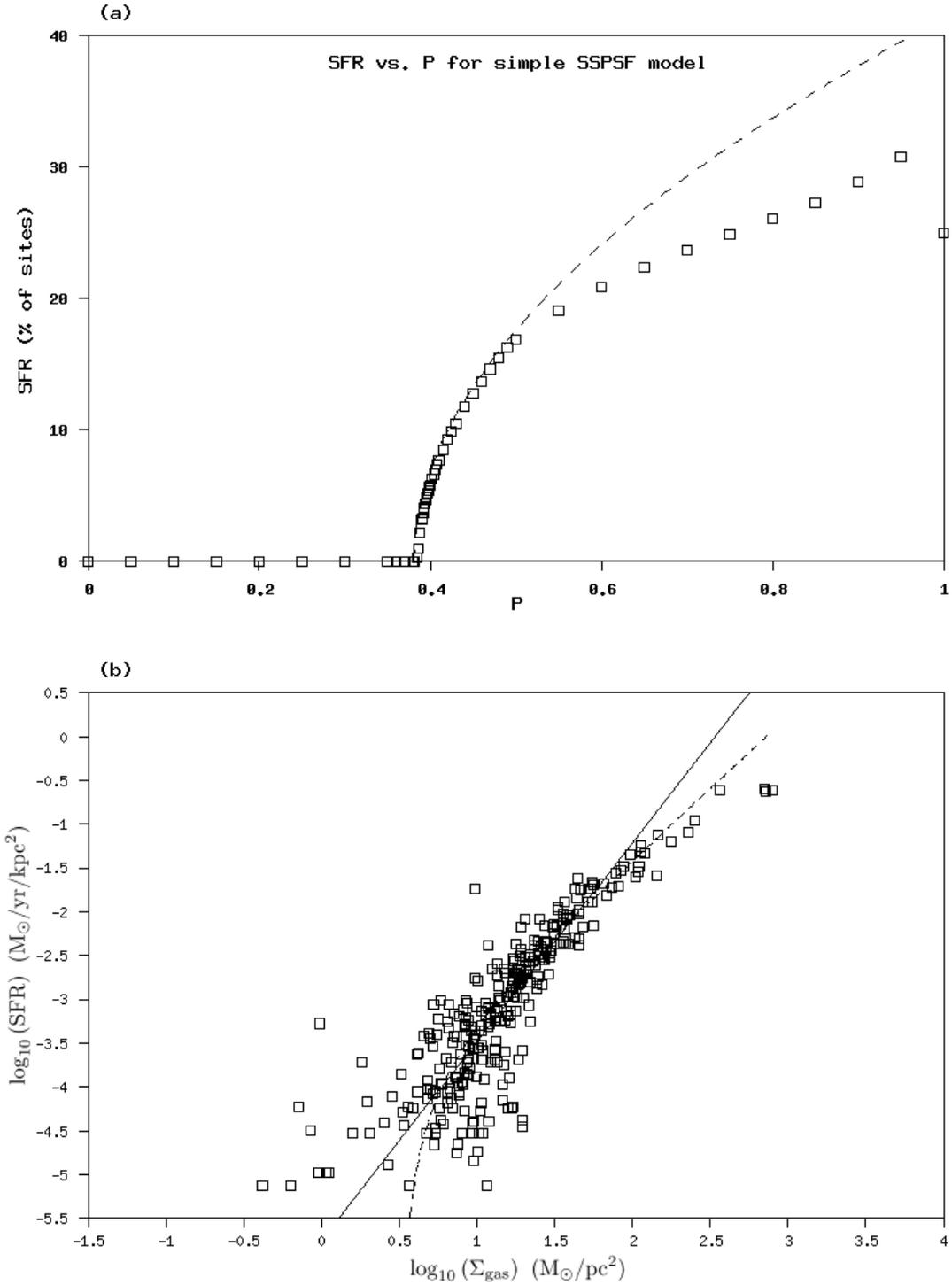}
\caption{(a) Results and fit of eq.~(3) for a simple SSPSF implementation described in the text. (b) Fits of the standard Schmidt law, eq.~(5),
and the modified form, eq.~(6) to numerical simulation results of Kravtsov (2003).}
\label{fig:theory}
\end{figure}

\begin{figure}[htp]
\centering
\epsfxsize=6in
\epsfbox{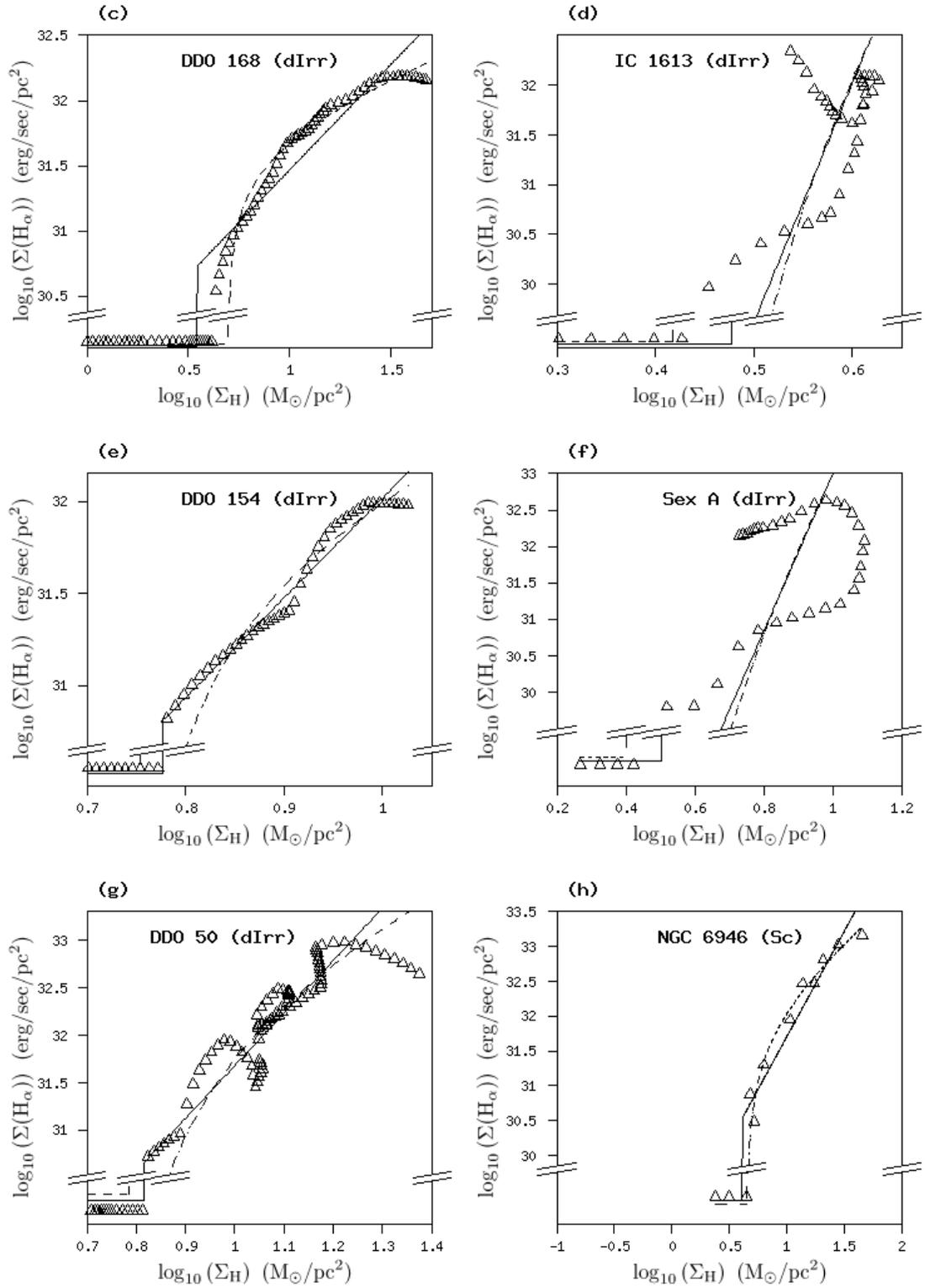}
\caption{Fits of standard and modified forms (eqns.~(5) and (6)) of Schmidt law to data compiled by Hunter et al (1998) (c-g) and Kennicutt (1989) (h-n).}
\label{fig:multi1}
\end{figure}

\begin{figure}[htp]
\centering
\epsfxsize=6in
\epsfbox{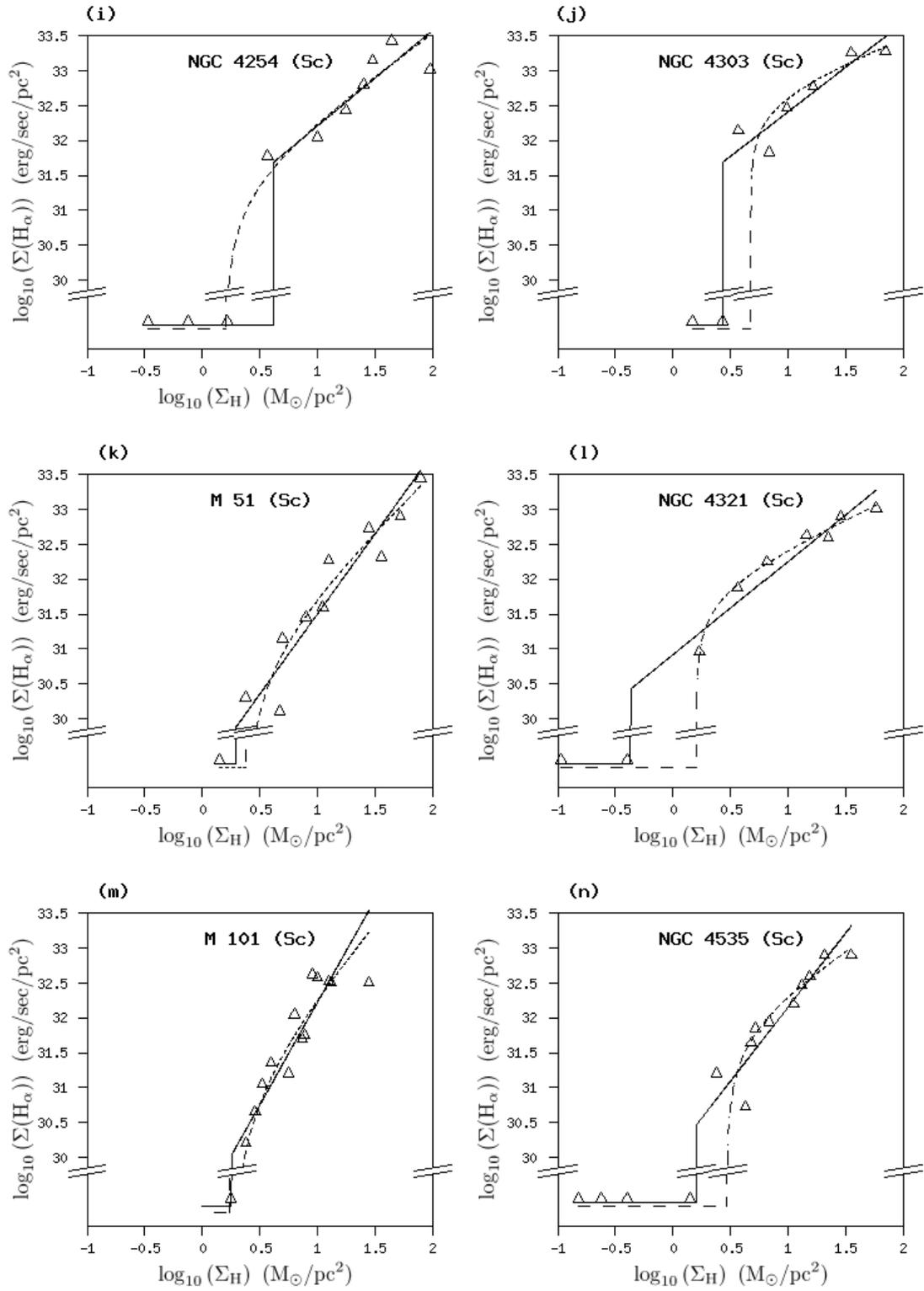}
\caption{Fits of standard and modified forms of Schmidt law, continued.}
\label{fig:multi2}
\end{figure}

\end{appendix}

%%%%%%%%%%%%%%%%%%%%%%%%%%%%%%%%%%%%%%%%%%%%%%%%%%%%%%%%%%%%%%%%%%%%%%%
%%%%%%%%%%%%%%%%%%%%%%%%%%%%%%%%%%%%%%%%%%%%%%%%%%%%%%%%%%%%%%%%%%%%%%%
\end{document}